\documentclass[10pt, a4paper, twocolumn]{article} 

\usepackage[square,numbers,sort&compress,comma]{natbib}

\usepackage{amsmath}
\usepackage{amssymb}
\usepackage{caption}
\usepackage{graphicx}
\usepackage{latexsym}
\usepackage{times}
\usepackage[hidelinks]{hyperref}
\usepackage[left=2.5cm,right=2.5cm, top=2cm, bottom=2cm]{geometry}
\usepackage{stfloats}
\usepackage{xcolor}
\usepackage{fancyhdr}
\usepackage{titlesec}

\fancypagestyle{plain}{
	\fancyhead[C]{\textit{\small Manuscript accepted by Proceedings of the Combustion Institute}}
	\fancyhead[R]{}
}

\renewenvironment{abstract}
{
	\begin{center}
		\textbf{Abstract}
	\end{center}
	\begin{quote}
		\setlength{\parskip}{0pt}
	}{
	\end{quote}           
}

\newenvironment{novelty}{
	\begin{center}
		\textbf{Novelty and significance statement}
	\end{center}
	\begin{quote}
		\setlength{\parskip}{0pt}
	}{
	\end{quote}
}

\titleformat{\section}
{\large\bfseries}
{\thesection}
{0.5em}
{}
\titleformat{\subsection}
{\normalsize\bfseries}
{\thesubsection}
{0.5em}
{}
\titlespacing*{\section}
{0pt}{10pt}{0pt}
\titlespacing*{\subsection}
{0pt}{10pt}{0pt}

\renewcommand{\captionsize}{\small}
\begin{document}

\setcounter{page}{1}
\title{\bf Large-eddy simulation of moderately dense evaporating sprays with particle-informed super-resolution}

\author{{\large Ruyue Cheng, Ali Shamooni, Andreas Kronenburg$^{*}$,}\\
	{\large Jan Wilhelm Gärtner, Thorsten Zirwes}\\[10pt]
	{\small Institute for Reactive Flows (IRST), University of Stuttgart, Pfaffenwaldring 31, 70569 Stuttgart, Germany}\\[-3pt]
	{\small *Corresponding author: kronenburg@irst.uni-stuttgart.de}}

\date{}

\linespread{1.03}
\baselineskip 12pt

\twocolumn[\begin{@twocolumnfalse}
	\maketitle
	
	\begin{center}
		\vspace{-10pt}
		\rule{0.92\textwidth}{0.5pt}
	\end{center}

\begin{abstract} 
In large-eddy simulation (LES) of dense sprays or sprays with pronounced clustering, evaporation rates can be inaccurate when the mesh is too coarse to provide realistic boundary conditions for the widely employed single droplet evaporation model. This is especially relevant to liquid spray combustion in practical applications. Deep learning-based super-resolution (SR) has recently emerged as a promising method for LES subgrid-scale modeling, capable of enhancing flow field resolution. This technique appears well-suited to reconstruct the local gas fields within the inter-droplet space that can be used to correct the evaporation rates. However, it has not yet been applied for this purpose. This paper presents an innovative SR approach -- particle-informed super-resolution (PISR) -- that approximates high-resolution flow fields for improved evaporation computation. It is validated with \textit{a priori}, \textit{a posteriori} and generalization tests on moderately dense sprays. The results show that PISR-LES can closely replicate the evaporation rates computed in a carrier-phase direct numerical simulation (CP-DNS), significantly reducing the discrepancy in the fuel mass fraction field between LES and CP-DNS. Furthermore, the PISR model exhibits robust generalization to cases unseen in training when varying air temperature, droplet diameter, turbulent Reynolds number and spray pattern.  
\end{abstract}

\begin{novelty}
This study is the first to apply super-resolution to the modeling of particle source terms in an LES. A novel SR approach and a tailored LES framework are presented to achieve a physically consistent Lagrangian particle advancement based on the super-resolved gas fields. Accurate modeling of evaporation from dense or clustered sprays in turbulent flows is crucial for practical combustion processes but remains challenging to date. The proposed approach provides a promising solution to this challenge. Moreover, it is applicable to any Euler-Lagrange simulations of particle-laden flows where high-resolution flow fields can improve the modeling of particle source terms.

\vspace{10pt}

\end{novelty}

\begin{center}
\vspace{-10pt}
\rule{0.92\textwidth}{0.5pt}
\end{center}
\vspace{5pt}

\end{@twocolumnfalse}] 

\section{Introduction\label{sec:introduction}} \addvspace{10pt}

In the vast majority of spray combustion simulations, droplets are not fully resolved but treated as Lagrangian particles. Their interactions with the gas phase are modeled within an Euler-Lagrange framework. Droplet evaporation models are generally based on the single droplet assumption, with heat and mass transfer driven by temperature and vapor concentration differences across a boundary layer between the droplet surface and the bulk flow~\cite{Faeth1977PiEaCS,Abramzon1989IJoHaMT}. However, the boundary layer around a droplet is typically not resolved in the simulations due to the prohibitive computational cost. Hence, the cell-averaged gas properties of the cell containing the droplet are commonly used as the boundary conditions imposed by the bulk flow. This leads to a grid dependence of evaporation rates when simulating sprays. For dilute and well dispersed sprays, the mesh should not be too fine, otherwise the cell-averaged values cannot reasonably represent the far-field bulk conditions~\cite{Sontheimer2021CaF}. Nevertheless, experiments have shown that in many practical combustion processes, droplets often burn as a cloud enveloped by a single flame (group combustion)~\cite{Nakabe1988CaF,Akamatsu1996SoC,Tsushima1998SoC,Mikami2005PotCI,MManish2021PotCI}. This combustion mode pertains not only to dense sprays but also to dilute sprays that exhibit clustering induced by turbulence or preferential flame propagation~\cite{CHIU1977CST,Akamatsu1996SoC,Pandurangan2024PotCI}. In this case, droplets in the interior of the cloud evaporate more slowly than those at the periphery, because their local ambient temperatures are lower and their local ambient vapor concentrations are higher due to evaporation from neighboring droplets~\cite{Sirignano1983PiEaCS,Sirignano2014PiEaCS}. Therefore, for droplets within the cloud, the boundary conditions for evaporation modeling should be based on the local ambient gas properties within the inter-droplet space, and finer meshes would be needed. A typical LES mesh is, however, too coarse to resolve this inter-droplet space and the cell-averaged gas properties cannot reasonably represent the local ambient conditions of droplets~\cite{Sirignano2014PiEaCS}.      

Over the past five years, deep learning-based super-resolution (SR) has emerged as a promising method for subgrid-scale (SGS) modeling in LES. Deep neural networks (the SR models) are trained to reconstruct high-resolution (HR) flow fields from low-resolution (LR) inputs. Thereby, the subgrid closures for the filtered governing equations can be more accurately estimated with the reconstructed HR fields. This method has been applied to the modeling of the SGS stresses, SGS mixing and chemical reaction~\cite{Bode2021PCI,Bode2022SIJA&CPiM,Bode2023PCI,Maejima2025JFM}. For LES of dense or clustered sprays, SR models can offer additional benefits: the reconstructed HR fields can approximate the local temperature and mass fraction profiles within the inter-droplet space and can be used for more precise evaporation modeling. However, to the best of the authors' knowledge, this additional capability of SR has not yet been explored.

\begin{figure*}[!b]
	\centering
	\includegraphics[width=\textwidth]{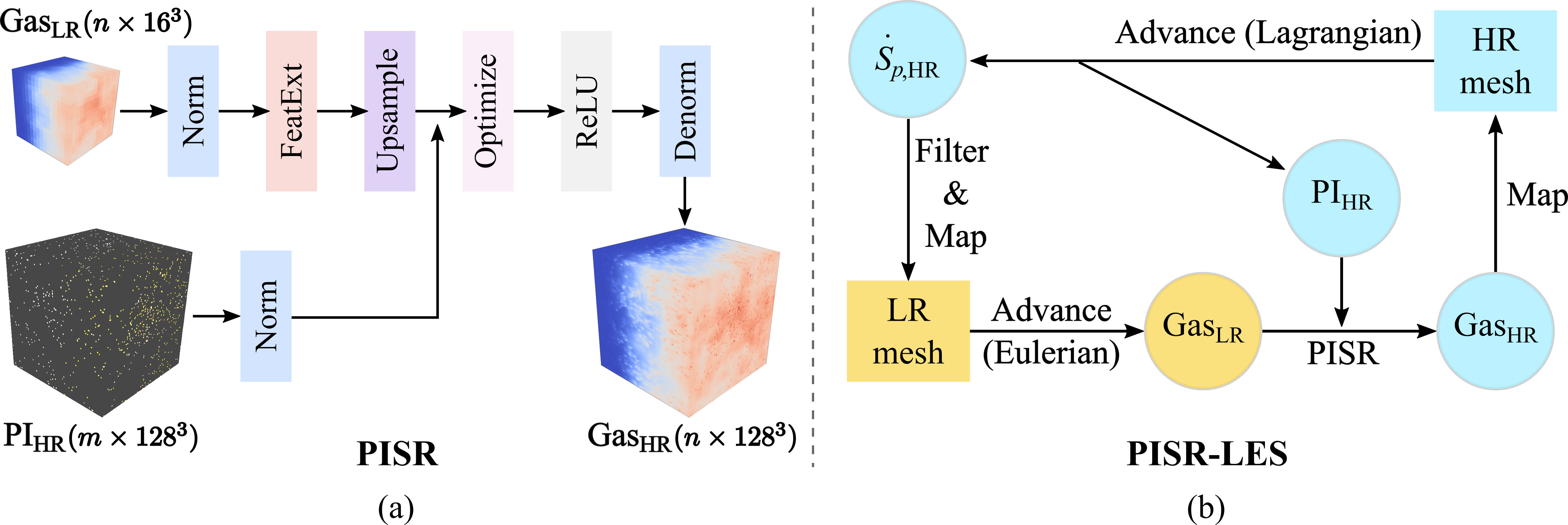}
	\vspace{-5 pt}
	\caption{\captionsize Workflow of the (a) PISR and (b) PISR-LES.}
	\label{fig:workflow}
\end{figure*}

There are challenges in applying SR to evaporation modeling, with the major one being the preservation of physical correlations between the Lagrangian particles and the reconstructed HR gas fields. The particle-induced local flow structures (e.g., vapor-rich pockets resulting from evaporation) are strongly correlated with particle states such as position, diameter and temperature. Therefore, it is difficult to reconstruct these local structures if feeding only LR gas fields to the SR model. However, feeding particle data to the SR model is also not straightforward. Particle data have to be transferred into cell-based particle information (e.g., number/mean diameter of particles per cell) in order to be processed together with the gas fields by the SR model. Recent \textit{a priori} SR studies for particle-laden flows have demonstrated that adding particle information from each LR cell to model input can improve the super-resolution of particle-induced SGS structures statistically~\cite{Shamooni2025PoF,Cheng2026AAS,Shamooni2025PotCI}. However, a statistical reconstruction of local flow structures is still inadequate for Lagrangian advancement, as the spatial correlation between the two phases is not guaranteed. 

In this paper, we propose a novel SR approach along with a novel LES implementation to address the challenges in SR for evaporation modeling. The methodology section clarifies the workflow and principles of the proposed approach, describes the simulation cases and solvers used to validate it and summarizes the training strategy. The validation results are presented in the results section, covering \textit{a priori}, \textit{a posteriori} and generalization tests, as well as computational performance.

\section{Methodology\label{sec:meth}} \addvspace{10pt}

\subsection{Particle-informed super-resolution (PISR)\label{subsec:SR}} \addvspace{10pt}

We propose particle-informed super-resolution (PISR). The key idea and novelty is that particle information from each high-resolution (HR) cell is used to guide the super-resolution of gaseous fields, which is obtained by advancing Lagrangian particles on an HR mesh. Figure~\ref{fig:workflow}(a) outlines the workflow of this SR approach. Low-resolution gas fields (Gas\textsubscript{LR}, e.g., $n$ fields at resolution $16^3$) are first normalized to the range [0,1]. The PISR model then extracts the shallow and deep features of Gas\textsubscript{LR}. With these features, the upsampling block preliminarily estimates the small-scale structures and super-resolves the gas fields to high resolution (e.g., $128^3$). Afterwards, HR particle information (PI\textsubscript{HR}), such as average particle diameter and temperature per HR cell, is fed to the model and concatenated with the gas fields. The subsequent model block utilizes this concatenated data to optimize the gas fields. A ReLU layer ($f(x)=max(0,x)$) at the end is applied to scalar fields to ensure non-negative values, acting as a physical constraint. All the gas fields are then denormalized to obtain the final output Gas\textsubscript{HR}. Each of the feature extraction and optimizing blocks consists mainly of a single Residual-in-Residual Dense Block (RRDB)~\cite{Wang2019CV-E2W}, an architecture well established for turbulence SR~\cite{Bode2021PCI,Nista2023PCI,Nista2024PRF}. Details of the upsampling block can be found in~\cite{Cheng2025PoF}.   

For this purpose, the LES case should contain two meshes: one coarse (LR) mesh for solving the LES governing equations of the gas phase, and one HR mesh for Lagrangian particle advancement. Figure~\ref{fig:workflow}(b) shows a schematic of the coupling between the two phases in PISR-LES: After each time step, the LR fields of all the gas properties required for Lagrangian advancement (Gas\textsubscript{LR}) are super-resolved by the trained PISR model to produce the HR fields (Gas\textsubscript{HR}). These HR gas fields will then be mapped to the HR mesh and used for particle advancement. Thereby, the mass, heat and momentum transfer between the two phases can be computed on the HR mesh with higher precision ($\dot{S}_{p,\text{HR}}$), and then filtered and mapped back to the LR mesh as source terms for the next-step LES solution of gas fields. The updated particle information (PI\textsubscript{HR}) will be used in the next-step super-resolution. Importantly, the additional HR mesh will not substantially increase the computational cost to the level of carrier-phase direct numerical simulation, because it is only used for Lagrangian advancement, whose computational cost scales with the number of tracked particles rather than mesh resolution.

By incorporating PI\textsubscript{HR} into model inputs, PISR enables the SR model to learn and preserve physical correlations between the particles and HR gas fields. This not only allows for a physically accurate reconstruction of the particle-induced local flow structures, but also facilitates the reconstruction of the larger SGS turbulence structures that influence particle displacement. SR models have long been trained to reconstruct SGS turbulence structures. However, they have struggled to fully meet the expectations due to the ill-posedness of turbulence SR. The main issue is that one input LR field can correspond to multiple HR solutions, owing to the random and chaotic nature of turbulence~\cite{Cheng2025PoF}. PISR largely changes the situation because HR gas fields are strongly correlated to PI\textsubscript{HR} in two-way coupled particle-laden flows. For example, even if only momentum interactions are considered, two-way coupled particles modulate turbulence across nearly all length scales, with the modulation pattern varying with the Stokes number~\cite{Ferrante2003PoF,Abdelsamie2012PoF}. Thereby, the additional model input PI\textsubscript{HR} considerably narrows the HR solution space and reduces the ill-posedness of turbulence SR. This eases the training and boosts the performance of SR models.    

\subsection{Simulation cases\label{subsec:valid}} 
\begin{figure}[h!]
	\centering
	\includegraphics[width=0.7\linewidth]{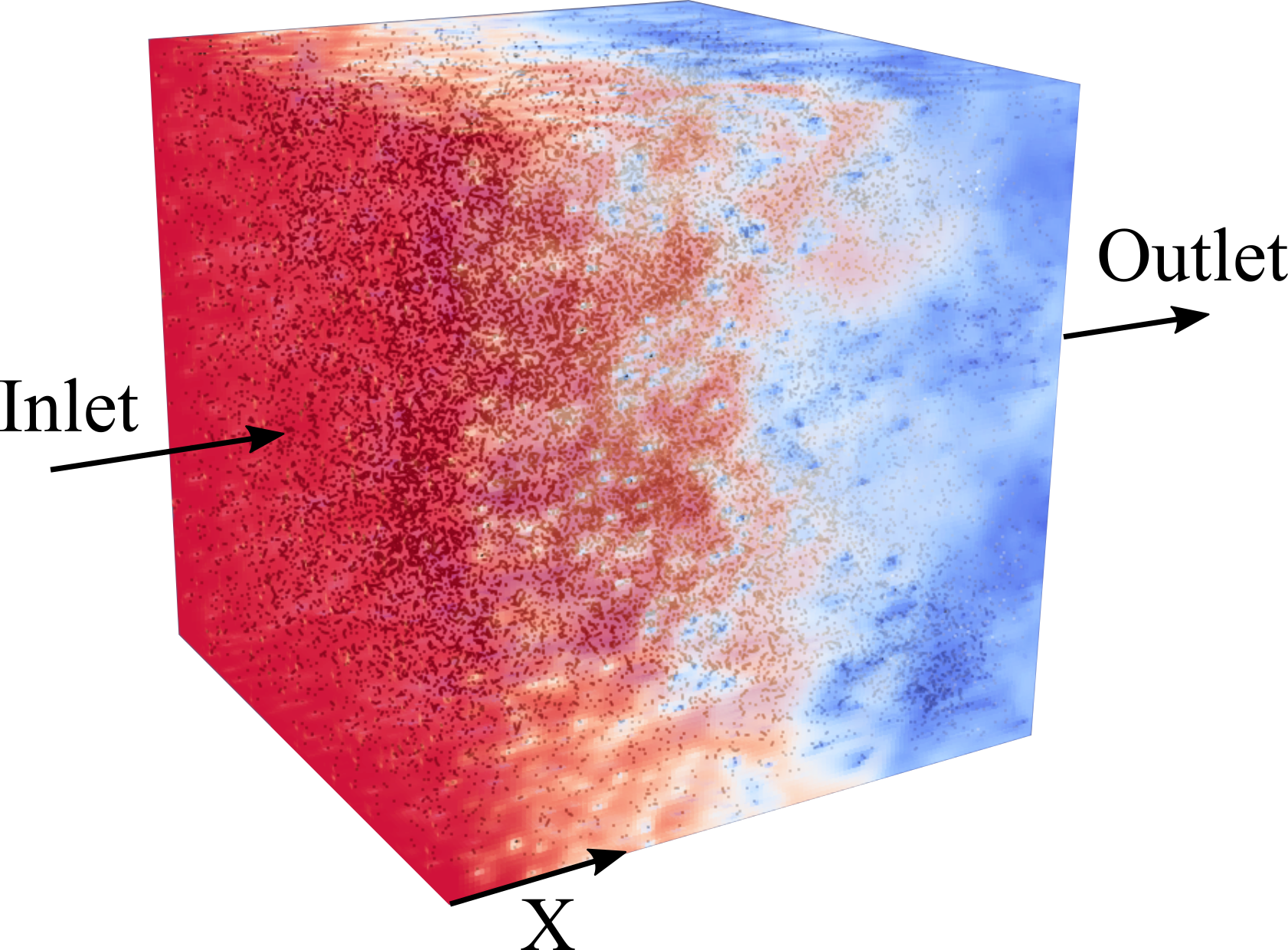}
	\vspace{8 pt}
	\caption{\captionsize Computational domain.}
	\label{fig:domain}
\end{figure}

\addvspace{10pt}
A carrier-phase direct numerical simulation (CP-DNS) of a moderately dense evaporating spray is conducted as a baseline case, following the setups in a previous study~\cite{Wang2019FTCa}. It is used to build the training dataset and to serve as the reference for the \textit{a priori} and \textit{a posteriori} tests of PISR. The computational domain is a cube with an inlet, outlet and four periodic boundaries, as shown in Fig.~\ref{fig:domain}. It contains a droplet-laden turbulent flow. The turbulence is homogeneous and isotropic at the inlet and decays downstream. Droplets (C\textsubscript{12}H\textsubscript{23}) are continuously injected from random positions at the inlet and initialized with the local flow velocity. A constant inlet liquid volume fraction ($\alpha_0$) is prescribed. Table~\ref{tab:inlet} summarizes the case setup of this baseline case. Some key parameters are changed for the generalization tests of PISR, with the specific changes detailed in Sec.~\ref{subsec:generalization}.
\begin{table}[ht!] \captionsize
	\caption{The setup of the baseline CP-DNS (subscript 0 indicates inlet condition).}
	\centerline{\begin{tabular}{cc}
			\hline 
			\rule{0pt}{1em}
			Parameter& Value     \\
			\hline
			\rule{0pt}{1em}
			Turbulent Reynolds number, $\mathrm{Re_{t,0}}$  & 40.3 \\
			Turbulence intensity, $U'_0/\overline{U}_0$ & 16.7\% \\
			Domain-to-integral-scale ratio, $L/L_{\mathrm{int,0}}$ & 4 \\
			Liquid volume fraction, $\alpha_0$ &  $8\times10^{-4}$ \\
			Air temperature, $T_0$ [K] & 2000 \\
			Droplet temperature, $T_{\mathrm{p,0}}$ [K] & 350 \\
			Droplet diameter, $d_0$ [$\mu$m] & 50 \\
			Grid-to-droplet size ratio, $\Delta x/d_0$ & 2 \\
			Turbulence resolution, $\Delta x/\eta_0$ & 0.5  \\
			Stokes number, $\mathrm{St_{\eta,0}}$ & 4.7  \\
			Average droplet spacing, $r_c/d_0$ & 8  \\	
			\hline
			\multicolumn{2}{l}{\footnotesize $\eta$: Kolmogorov length scale}\\ 
	\end{tabular}}
	\label{tab:inlet}
	\vspace{-10 pt}
\end{table}

In the \textit{a posteriori} tests, the PISR model is coupled with an LES solver to conduct PISR-LES. The results are compared with standard LES (i.e., LES without PISR) and CP-DNS. The mesh sizes of the CP-DNS and LES are $128^3$ and $16^3$ respectively. Therefore, in the PISR-LES, the filtered conservation equations are solved on an LR mesh of size $16^3$ and the Lagrangian particles are advanced on an HR mesh of size $128^3$. The inlet air velocities of the LES are sampled and filtered from the CP-DNS over time. The Courant number is kept below 0.3. All the solvers are developed based on the \textit{sprayFoam} solver from OpenFOAM-v2406. All the simulations are conducted in an Euler-Lagrange framework with a point-particle assumption. The two-way coupling is modeled with a particle-source-in-cell (PSI-cell) approach~\cite{Crowe1977JoFE}. The equations for Lagrangian particle tracking are: 
\begin{equation}
	\frac{\mathrm{d}\boldsymbol{x}_{\mathrm{p}}}{\mathrm{d}t} = \boldsymbol{u}_{\mathrm{p}},
\end{equation}
\begin{equation}
	\frac{\mathrm{d}\boldsymbol{u}_{\mathrm{p}}}{\mathrm{d}t} = \frac{3\mu_\mathrm{c} \mathrm{C}_\mathrm{D}\mathrm{Re}_{\mathrm{p}}}{4\rho_{\mathrm{p}} d^2}(\boldsymbol{u}_\mathrm{c}(\boldsymbol{x}_{\mathrm{p}})-\boldsymbol{u}_{\mathrm{p}}),
\end{equation}
where $\mu_\mathrm{c}$, $\mathrm{C}_{\mathrm{D}}$, $\mathrm{Re}_{\mathrm{p}}$, $\rho_{\mathrm{p}}$ and $d$ are carrier phase dynamic viscosity, particle drag coefficient, particle Reynolds number, particle density and particle diameter, respectively. The term $\boldsymbol{u}_\mathrm{c}(\boldsymbol{x}_{\mathrm{p}})$ is the carrier phase velocity interpolated at the particle position. Particle collisions are neglected. In the standard LES, $\boldsymbol{u}_\mathrm{c}(\boldsymbol{x}_{\mathrm{p}})$ is interpolated from the filtered velocity. The equations governing droplet evaporation and temperature are:
\begin{equation}
	\dot{m}_{\mathrm{p}} = - \pi d \rho_{\mathrm{v,s}} \mathrm{Sh} D \mathrm{ln}(1+\frac{X_{\mathrm{v,s}}-X_{\mathrm{v,ab}}}{1-X_{\mathrm{v,s}}}),
\end{equation}
\begin{equation}
	m_{\mathrm{p}} C_{p,\mathrm{p}}\frac{\mathrm{d}T_{\mathrm{p}}}{\mathrm{d}t} = h A_{\mathrm{p}}(T_{\mathrm{ab}}-T_{\mathrm{p}}) + \dot{m}_{\mathrm{p}} L_{v},
\end{equation}
where $\mathrm{Sh}$, $D$, $X$, $h$, $A_{\mathrm{p}}$ and $L_{v}$ are Sherwood number, diffusion coefficient, molar fraction, heat transfer coefficient, droplet surface area and the latent heat of evaporation, respectively. Subscripts $\mathrm{v}$, $\mathrm{s}$ and $\mathrm{ab}$ denote vapor, droplet surface and ambient gas condition, respectively. The volume-averaged values of the cell containing the droplets are used directly as $X_{\mathrm{v,ab}}$ and $T_{\mathrm{ab}}$. For more details of the momentum, heat and mass transfer modeling, the reader is referred to the \textit{sphereDrag}, \textit{RanzMarshall} and \textit{liquidEvaporationBoil} models in OpenFOAM. Due to the mesh dependency of $X_{\mathrm{v,ab}}$ and $T_{\mathrm{ab}}$, evaporation rates differ significantly between standard LES and CP-DNS, which enables a clear demonstration of the capability of PISR. In both standard LES and PISR-LES, the subgrid turbulence is modeled with OpenFOAM's Smagorinsky model. In principle, PISR could also be used for subgrid turbulence modeling, but this can lead to additional differences between the two LES. It is therefore not considered in the current study as the focus is on the model performance in evaporation modeling.   

\subsection{Training the PISR model\label{subsec:training}} \addvspace{10pt}
The training data are collected from the baseline case after the simulation reaches a statistically steady state: the time-averaged outlet mass flux converges, with no notable change in the time-averaged streamwise profile of volumetric evaporation rate. 650 snapshots of the flow are sampled with a time interval of approximately $0.4t_{\mathrm{ft}}$, where $t_{\mathrm{ft}}$ is the flow-through time. In the current study, Gas\textsubscript{HR} comprises seven fields: density, pressure, mass fraction of C\textsubscript{12}H\textsubscript{23}, temperature, and three velocity components; PI\textsubscript{HR} also comprises seven fields: volume fraction of particles in each cell, Sauter mean diameter of the particles in each cell, average particle temperature, velocities and age (i.e., the physical time elapsed since injection), weighted by $d^2$. Therefore, each snapshot is of size $14\times 128^3$. During training, these snapshots are randomly cropped into smaller cubes of size $14\times 64^3$, which is a typical technique for data augmentation. The seven gas fields are then extracted as the ground-truth reference (Gas\textsubscript{GT}) for the model output and the remaining PI\textsubscript{HR} of size $7\times 64^3$ becomes part of the model input. The other part of the model input, Gas\textsubscript{LR} of size $7\times 8^3$, is generated by Favre filtering and downsampling Gas\textsubscript{GT} with a filter size of $8^3$. This means that the PISR model is trained to achieve an SR factor of eight.

The model is implemented and trained using a customized BasicSR framework~\cite{basicsr} in PyTorch. The training runs for 100,000 iterations until the loss converges, with a batch size of 32 and a learning rate of $2\times 10^{-4}$. The loss function contains pointwise losses, which calculate the mean absolute errors in the seven gas fields and their gradients between the model output and Gas\textsubscript{GT}, viz.
\begin{align}\label{eq:loss}
	L &= 0.5L_{\text{pixel}}+0.5L_{\text{gradient}} \notag\\
	&= 0.5\times \langle{|\text{Gas}_{\text{HR}} - \text{Gas}_{\text{GT}}|}\rangle \notag\\ 
	&\hspace{0.5cm} + 0.5\times \langle{|\nabla \text{Gas}_{\text{HR}} - \nabla \text{Gas}_{\text{GT}}|}\rangle,
\end{align}
where $\langle .\rangle$ denotes an average across all data points and fields of a training batch. Pointwise losses drive the super-resolution toward optimal overall spatial fidelity and are fundamental loss components in turbulence SR. The adversarial or spectral loss component, which has been applied in previous studies to address the limitation of pointwise losses in turbulence SR (e.g.,~\cite{Bode2021PCI,Nista2023PCI,Tofighian2024PoF,Cheng2025PoF}), is not employed. This is because they are essentially statistical constraints and can compromise spatial fidelity while improving the statistical accuracy~\cite{Cheng2025PoF}. For example, subgrid eddies are reconstructed with correct energies but appear at random positions. This might not be a concern when using SR to calculate SGS stresses/fluxes, because these subgrid closures account for the statistical effects of the subgrid scales. However, this can lead to unphysical local flow fields surrounding the particles and unreliable Lagrangian advancement. Moreover, the limitation of pointwise losses becomes less significant with PISR due to the additional input data PI\textsubscript{HR}. Therefore, the model trained with Eq.~\ref{eq:loss} can already produce fairly good results, which will be illustrated and discussed in the results section. 

\section{Results\label{sec:results}} \addvspace{10pt}
\subsection{\textit{A priori} test\label{subsec:apriori}} \addvspace{10pt}
\begin{figure*}[!t]
	\centering
	\includegraphics[width=0.9\textwidth]{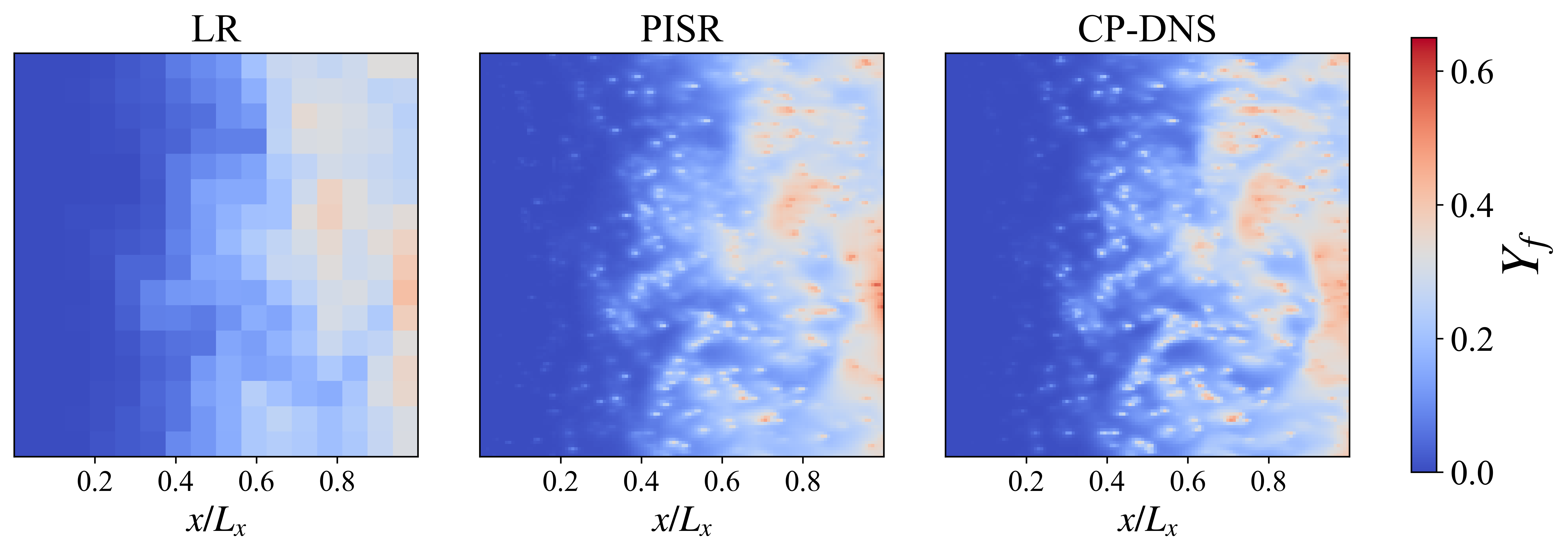}
	\caption{\captionsize \textit{A priori} test: fuel mass fraction in a longitudinal section of a model input (LR), output (PISR) and the target (CP-DNS).}
	\label{fig:apriori-Ycontour}
	\vspace{-5 pt}
\end{figure*}
\begin{figure*}[!t]
	\centering
	\includegraphics[width=0.95\textwidth]{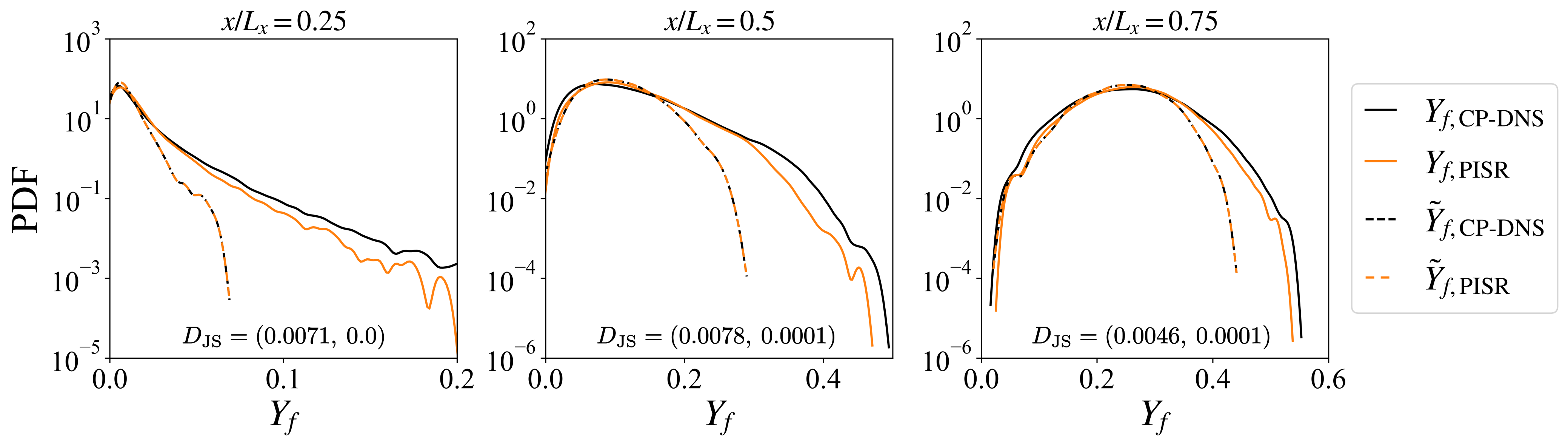}
	\caption{\captionsize \textit{A priori} test: PDFs of fuel mass fraction in three transverse sections. The two values of $D_{\mathrm{JS}}$ quantify the differences in the $Y_f$ (first value) and $\tilde{Y}_f$ (second value) distributions.}
	\label{fig:apriori-YPDF}
	\vspace{-5 pt}
\end{figure*}
After training, an \textit{a priori} test is first conducted for the PISR model. Twenty snapshots of the baseline CP-DNS are sampled over $15t_{\mathrm{ft}}$ as the test dataset. The sampling window is separated from the training data by $4t_{\mathrm{ft}}$ to ensure decorrelated test and training data. The flow fields of these snapshots are Favre-filtered and downsampled to LES resolution and fed to the model with their corresponding PI\textsubscript{HR}. Figure~\ref{fig:apriori-Ycontour} qualitatively illustrates the model's performance in reconstructing the fuel mass fraction field ($Y_f$) for one of the test samples. It can be observed that the $Y_f$ field of the CP-DNS features many small-scale structures resulting from droplet evaporation. These fine structures are lost in the LR input but effectively reconstructed by the PISR model. Clearly, the model output exhibits strong similarity to the CP-DNS reference. The normalized root-mean-squared error (NRMSE, defined as in \cite{Cheng2025PoF}) of $Y_f$ averaged over the whole test dataset is $0.09$. Figure~\ref{fig:apriori-YPDF} shows the probability density functions (PDF) of $Y_f$ and its Favre-filtered value $\tilde{Y}_{f}$ at three streamwise locations, computed by pooling data from the whole test dataset. The Jensen-Shannon divergence ($D_{\mathrm{JS}}$)~\cite{Lin1991ITIT} reported in Fig.~\ref{fig:apriori-YPDF} quantifies the difference between the probability distributions of the PISR output and CP-DNS, which is defined as  
\begin{align}\label{eq:Djs}
	D_{\mathrm{JS}}(p||q) = & 0.5\sum\limits_{i} p_i\log_{2}(\frac{2p_i}{p_i+q_i}) \notag\\
	& + 0.5\sum\limits_{i} q_i\log_{2}(\frac{2q_i}{p_i+q_i}),
\end{align}
 where $p$ and $q$ are the two discrete probability distributions being compared ($\sum\limits_{i} p_i = \sum\limits_{i} q_i = 1$). This divergence is bounded between 0 and 1, with smaller values indicating smaller differences (0: exact overlap, 1: no overlap). The bin width for constructing $p$ and $q$ is set to $1/40$ of the range of the reference distribution (i.e., CP-DNS for HR data and filtered CP-DNS for LR data) throughout this paper. Although the PISR model tends to slightly underestimate $Y_f$ at larger values, the correspondence between the model output and CP-DNS is good overall. The PDFs of $\tilde{Y}_{f}$, in particular, overlap perfectly, which indicates an excellent statistical cycle-consistency (i.e., $\widetilde{\text{Gas}}$\textsubscript{HR} $=$ Gas\textsubscript{LR}) that is not inherently guaranteed by SR models without special design~\cite{Kim2021JFM,Shamooni2025PoF}. 

\subsection{\textit{A posteriori} test\label{subsec:aposteriori}} \addvspace{10pt}
\begin{figure}[!b]
	\centering
	\vspace{-10 pt}
	\includegraphics[width=0.8\linewidth]{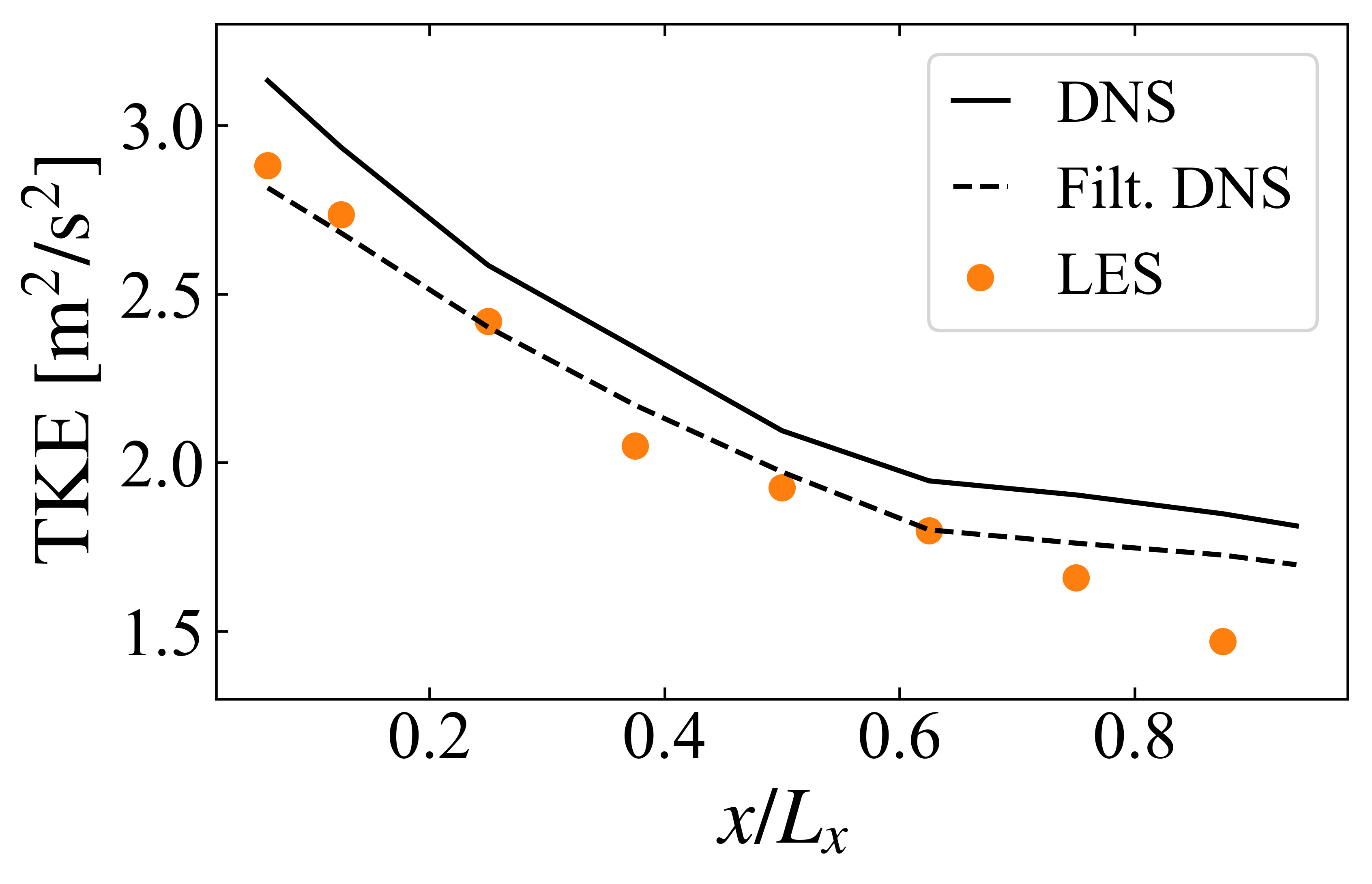}
	\caption{\captionsize Decay of turbulent kinetic energy along the streamwise direction (droplet-free).}
	\label{fig:TKE}
\end{figure}
When a PISR model is deployed in LES, the model inputs (i.e., Gas\textsubscript{LR} and PI\textsubscript{HR}) are no longer derived directly from CP-DNS. They contain errors from SGS models, which accumulate over time. As a result, the performance of the PISR model can differ from that observed in the \textit{a priori} test. Therefore, a PISR-LES of the baseline case is conducted and compared with a standard LES and CP-DNS to validate the PISR approach. All the results presented from this subsection onward are time-averaged over $15t_{\mathrm{ft}}$ with 20 samples after the simulation reaches a statistically steady state, unless otherwise specified. Since the inlet air velocities of the LES are filtered from the CP-DNS, the sampling windows for this \textit{a posteriori} test and for training data collection are separated by more than $20t_{\mathrm{ft}}$, with the training data sampling starting later. This ensures that the test data are not seen by the PISR model in training. 

\begin{figure}[!t]
	\centering
	\includegraphics[width=\linewidth]{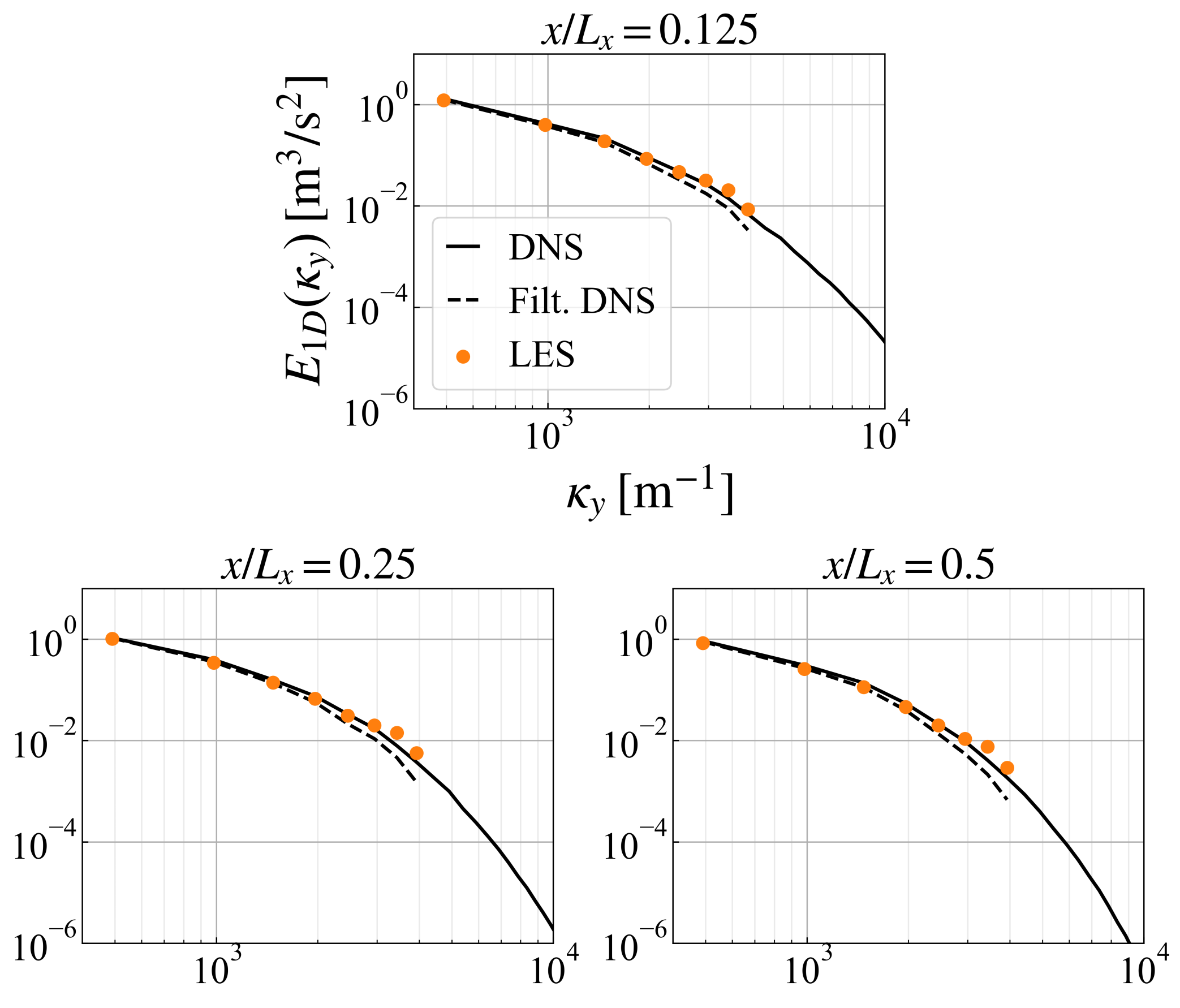}
	\caption{\captionsize One-dimensional velocity spectra in three transverse sections (droplet-free). $\kappa_y$ denotes the wavenumber in the y-direction.}
	\label{fig:spectra}
	\vspace{-5 pt}
\end{figure}

Preliminary simulations without droplets verify that the LES adequately approximates the DNS flow fields. As evidenced by Fig.~\ref{fig:TKE}, the decay of the turbulent kinetic energy (TKE) in LES matches the results of the filtered DNS (filtered after simulation) well from the inlet to about three quarters of the domain length ($L_x$). Good agreement is also observed in the one-dimensional velocity spectra (Fig.~\ref{fig:spectra}, defined as in~\cite{Cheng2025PoF}). Such good agreement in turbulence statistics is needed, because now any notable deviation between the droplet-laden LES and filtered CP-DNS upstream of $x/L_x = 0.75$ can be associated with the Lagrangian particle advancement. The effects of the PISR can be directly assessed.

\begin{figure}[!h]
	\centering
	\includegraphics[width=0.9\linewidth]{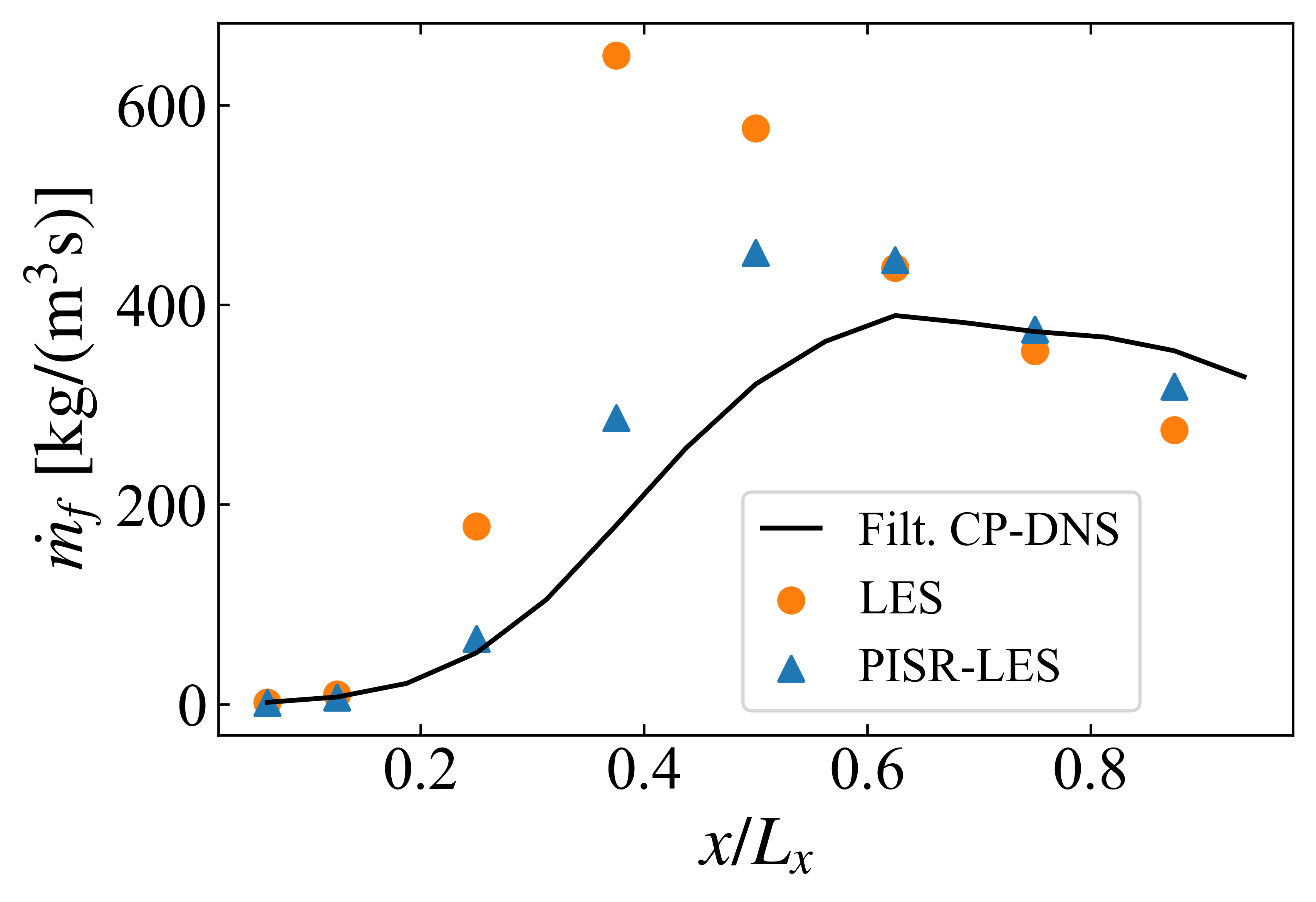}
	\caption{\captionsize Plane-averaged volumetric evaporation rates.}
	\label{fig:apost-mf}
\end{figure}

\begin{figure}[!h]
	\centering
	\includegraphics[width=0.9\linewidth]{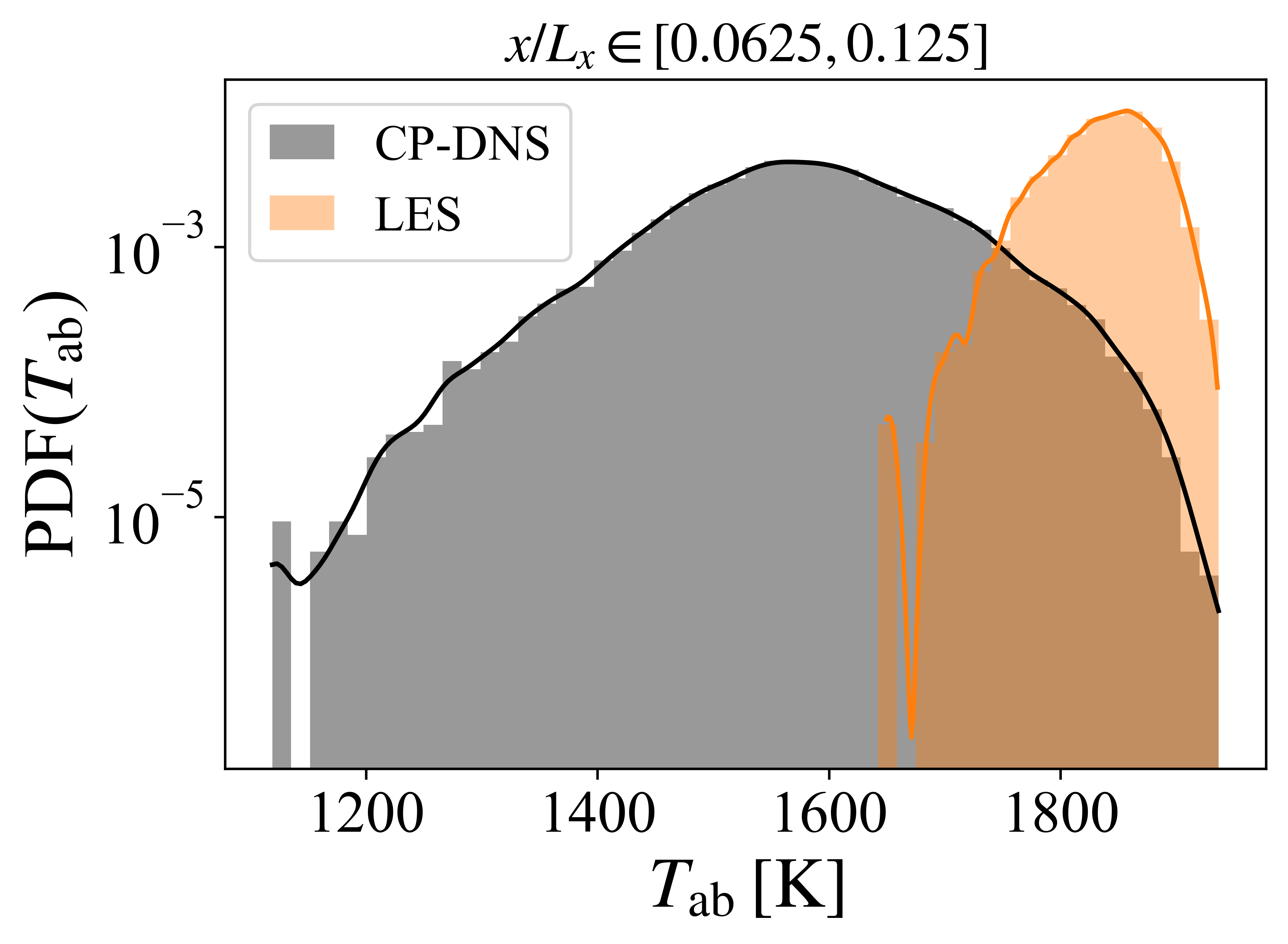}
	\caption{\captionsize PDFs of the ambient gas temperatures seen by droplets.}
	\label{fig:apost-Tab}
	\vspace{-5 pt}
\end{figure}

\begin{figure}[!h]
	\centering
	\vspace{-5 pt}
	\includegraphics[width=\linewidth]{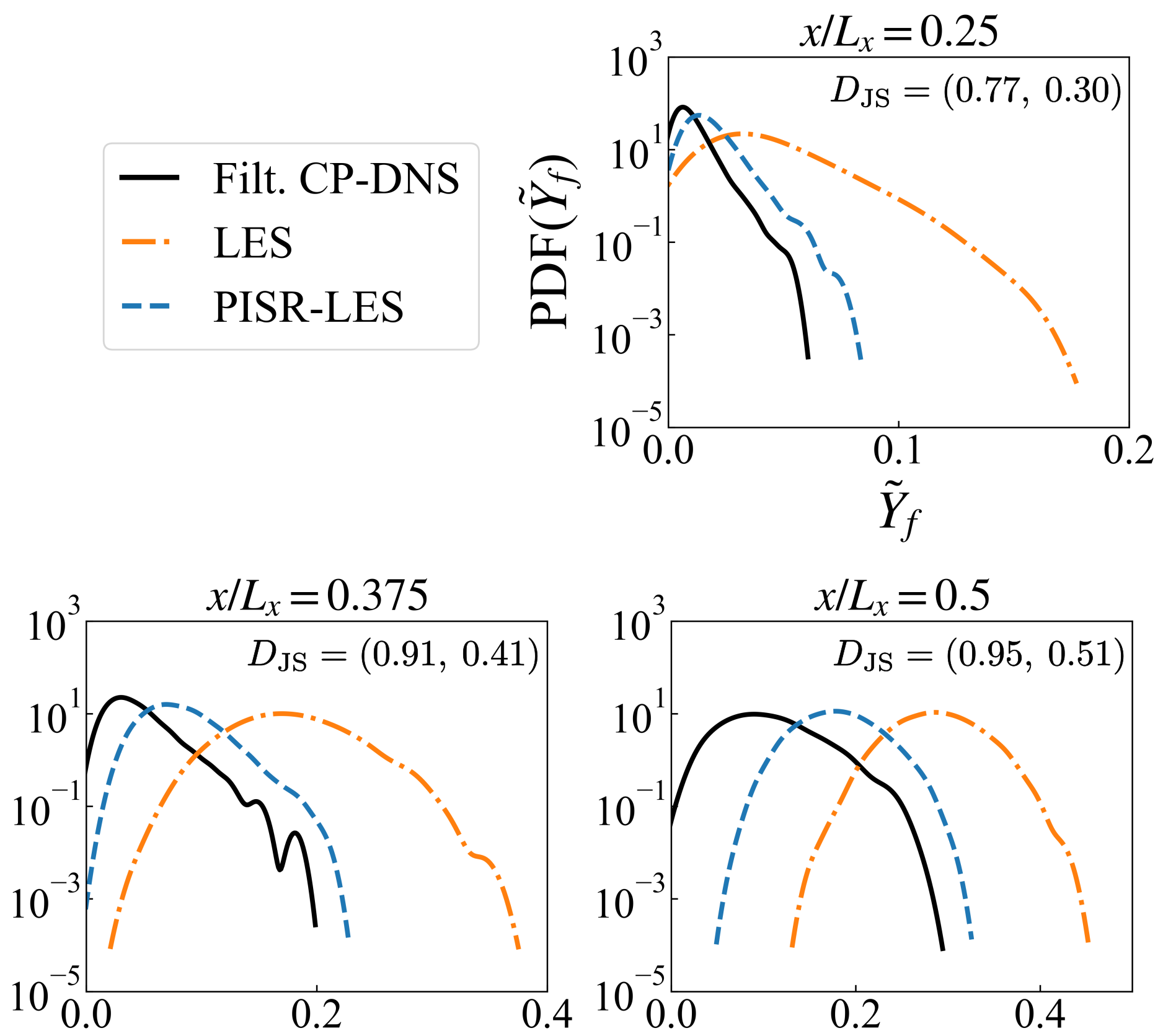}
	\vspace{-5 pt}
	\caption{\captionsize PDFs of filtered fuel mass fraction in three transverse sections. $D_{\mathrm{JS}}$ values are reported as (LES, PISR-LES).}
	\label{fig:apost-YPDF}
\end{figure}

The remainder of this subsection presents the droplet-laden results. The CP-DNS is filtered where necessary to provide a comparable reference for the standard LES and PISR-LES. Figure~\ref{fig:apost-mf} compares the streamwise profile of volumetric evaporation rates. A relatively large discrepancy can be observed in the evaporation profile between the standard LES and filtered CP-DNS. This is because droplets in the two simulations perceive different ambient conditions. As shown in Fig.~\ref{fig:apost-Tab}, the CP-DNS captures the coupled local cooling induced by neighboring droplets, leading to a broader PDF of droplet ambient temperature ($T_{\mathrm{ab}}$). In contrast, the standard LES neglects group evaporation effects and exposes droplets to higher ambient temperatures averaged over a larger region. Therefore, in the standard LES, droplets heat up faster and exhibit higher evaporation rates. Pronounced discrepancies between the standard LES and filtered CP-DNS are also evident in the fuel mass fraction fields (Fig.~\ref{fig:apost-YPDF}). The wide-spread PDFs of $\tilde{Y}_{f}$ are mainly caused by droplet clustering. For example, at $x/L_x = 0.25$, the number of droplets per cell in the filtered CP-DNS ranges from 0 to 31. A positive correlation is observed between droplet number density and $\tilde{Y}_{f}$. The higher evaporation rates in the standard LES lead to wider PDFs. The PISR model successfully approximates the flow fields at CP-DNS resolution, thus the results of PISR-LES show significantly improved agreement with the filtered CP-DNS.
\begin{figure}[!t]
	\centering
	\includegraphics[width=1\linewidth]{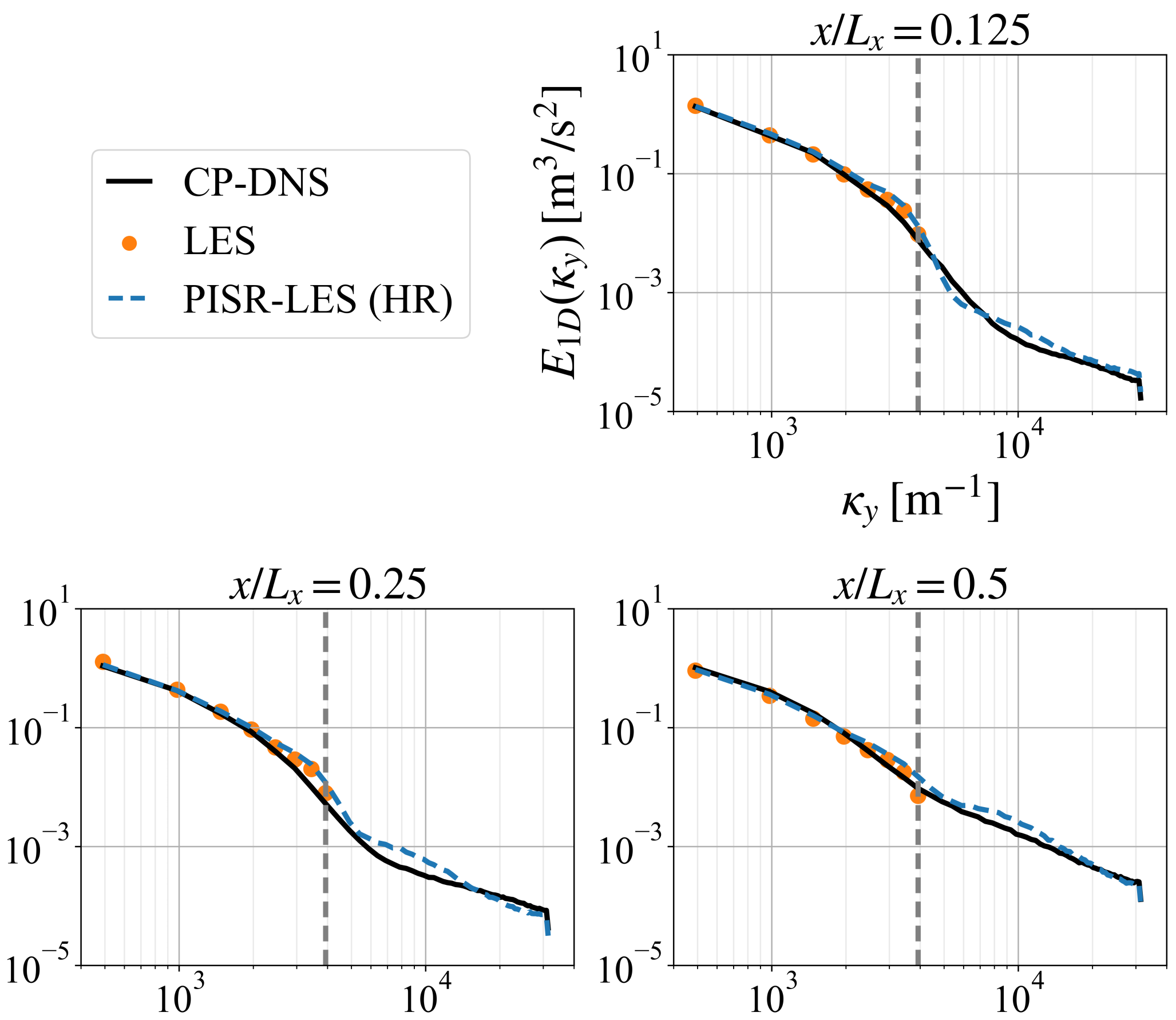}
	\vspace{-10 pt}
	\caption{\captionsize One-dimensional velocity spectra in three transverse sections.}
	\label{fig:apost-E1Dy}
\end{figure}

The one-dimensional velocity spectra of the flow fields reconstructed on the HR mesh in the PISR-LES are compared with the CP-DNS results in Fig.~\ref{fig:apost-E1Dy}. The vertical dashed gray line marks the cutoff wavenumber (i.e., highest resolved wavenumber on the LR mesh and in the standard LES). The energy at the wavenumbers immediately below this cutoff ($2\times 10^3 <\kappa_y< 4\times 10^3$) is slightly overestimated by PISR, which is attributed to the error in turbulence modeling. As illustrated by the comparison of droplet-free DNS and LES (Fig.~\ref{fig:spectra}), the energy at these wavenumbers in the LES is slightly higher than that of the filtered DNS. What is important is that in Fig.~\ref{fig:apost-E1Dy} the subgrid part of the spectra (i.e., for wavenumbers higher than the cutoff) are reconstructed by the PISR model fairly well. Significant energy underestimation in the subgrid spectra is often reported for SR models trained with only pointwise losses, which is due to the ill-posedness of turbulence SR~\cite{Kim2021JFM,Cheng2025PoF,Cheng2026AAS}. However, this does not appear in Fig.~\ref{fig:apost-E1Dy}, which supports the claim in Sec.~\ref{subsec:SR} that PISR largely reduces the ill-posedness and will boost model performance.   

\subsection{Generalization test\label{subsec:generalization}} \addvspace{10pt}
Generalization capability is a highly desirable property of machine learning models for real-world applications. Therefore, the \textit{a posteriori} test is extended to five additional cases (Table~\ref{tab:gCases}) that are not used for model training, with variations in inlet temperature (T1600), droplet diameter (D40), turbulent Reynolds number (Ret84) and spray pattern (Dpoly and HCone). Case Dpoly considers a polydisperse spray, where the inlet droplet diameters are randomly sampled from a Rosin-Rammler distribution that peaks between $50$ and $55\,\mu\mathrm{m}$. A spread factor of 3 is adopted, which is a common value for practical sprays~\cite{Lefebvre2017}. Case HCone features a hollow-cone spray injected from the inlet center, with inner and outer half-cone angles of $25^\circ$ and $40^\circ$ at injection. A lower inlet air temperature is prescribed to move toward more realistic injection conditions. The domain and mesh sizes in all the generalization cases are the same as the baseline case. PISR does not require a fixed mesh size. Generalization across different mesh sizes is implicitly achieved in all the test cases, as the model is trained with cropped flow samples (see Sec.~\ref{subsec:training}). 
\begin{table*}[!b] \captionsize
	\caption{CP-DNS setup for generalization tests.}
	\centerline{\begin{tabular*}{0.9\linewidth}{@{\extracolsep{\fill}} c c c c c c c c}
			\hline 
			\rule{0pt}{1em}
			Case & $\mathrm{Re_{t,0}}$ & $T_0$ & $d_0$ & $\Delta x/d_0$ & $\Delta x/\eta_0$ & $\mathrm{St_{\eta,0}}$ & $r_c/d_0$ \\
			\hline
			\rule{0pt}{1em}
			T1600 & 57.5 & 1600 & - & - & 0.7 & 6.4 & - \\
			D40 &  - & - & 40 & 2.5 & - & 3.0 & -\\
			Ret84 & 83.9 & - & - & - & 0.9 & 14.2 & - \\
			Dpoly &  - & - & 20-80 & 1.25-5 & - & 0.8-12.1 & 5-22\\
			HCone  &  123 & 1000 & - & - & 1.2 & 12.6 & 1-8$^{\dagger}$\\
			\hline 
			\multicolumn{8}{l}{\footnotesize --: same as baseline; $\dagger$: local values within the spray cone region.}\\
	\end{tabular*}}
	\label{tab:gCases}
\end{table*}
\begin{figure}[!t]
	\centering
	\includegraphics[width=\linewidth]{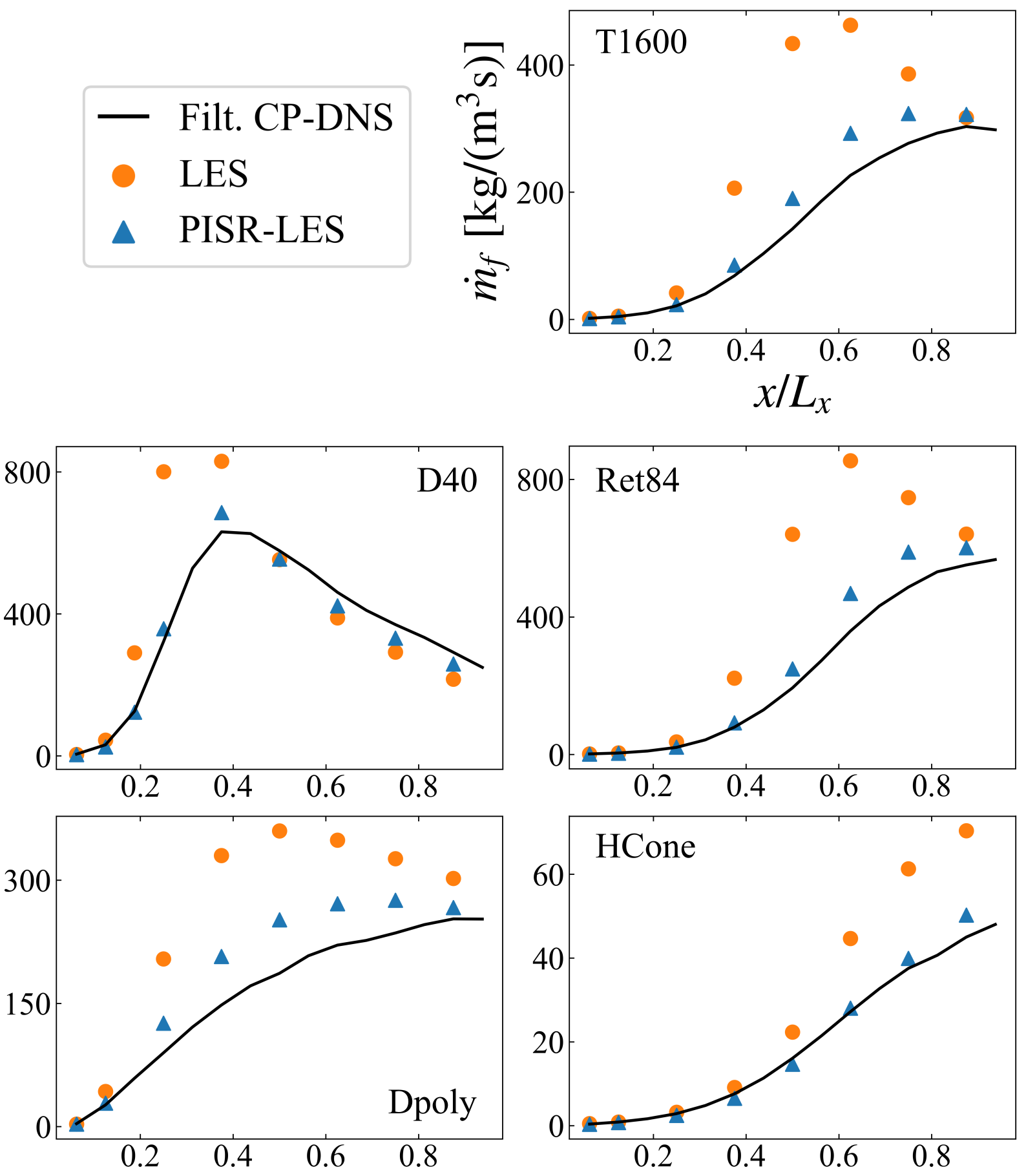}
	\caption{\captionsize Plane-averaged volumetric evaporation rates of generalization test cases.}
	\label{fig:gen-mf}
\end{figure}
\begin{figure}[!t]
	\centering
	\includegraphics[width=\linewidth]{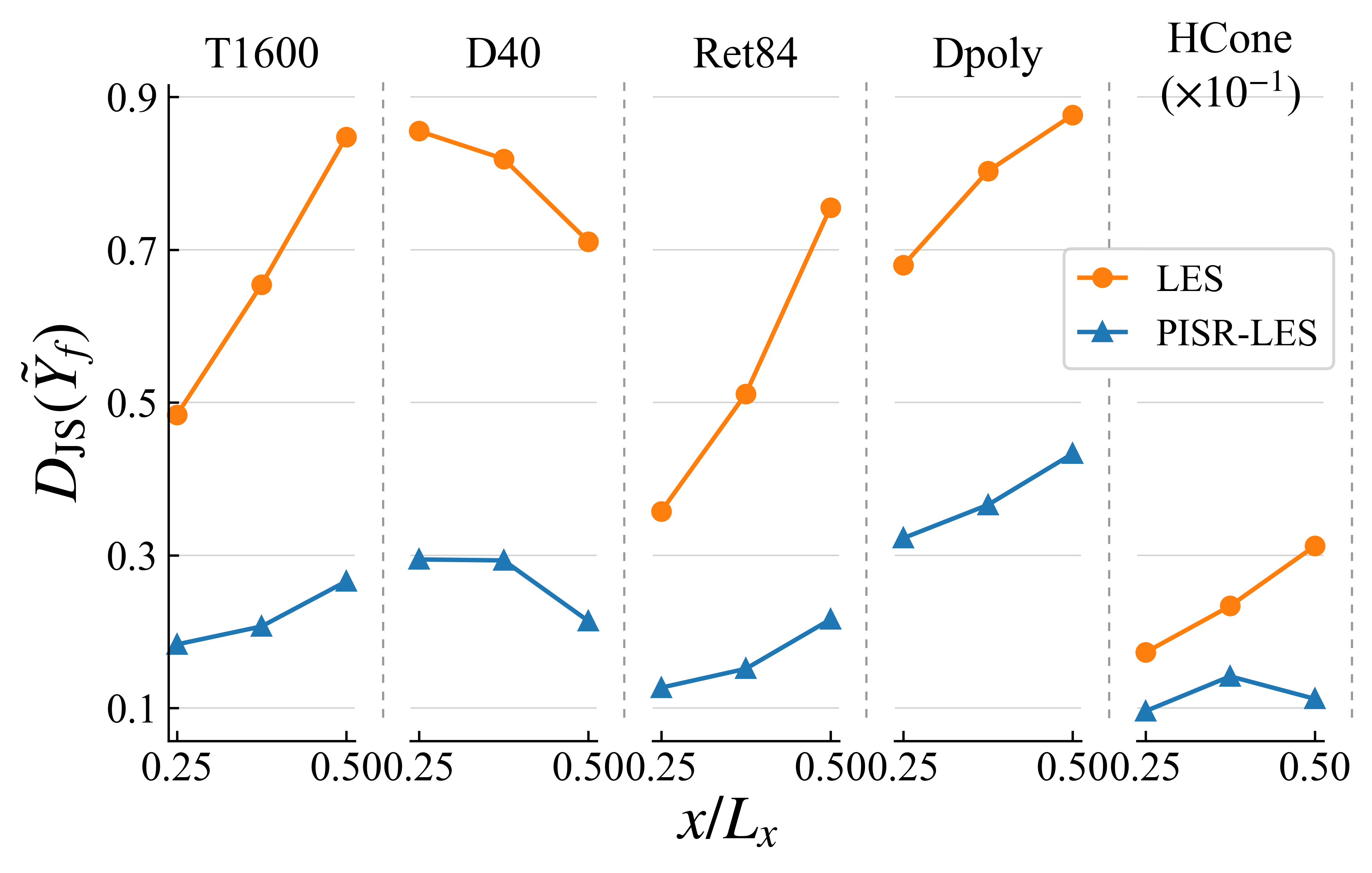}
	\caption{\captionsize  Quantified deviations (Eq.~\ref{eq:Djs}) of the predicted $\tilde{Y}_{f}$ distributions from the filtered CP-DNS results in three transverse sections for five test cases. Smaller $D_{\mathrm{JS}}$ values indicate better agreement with the filtered CP-DNS.}
	\label{fig:gen-YPDF}
\end{figure}
\begin{figure}[!t]
	\centering
	\includegraphics[width=\linewidth]{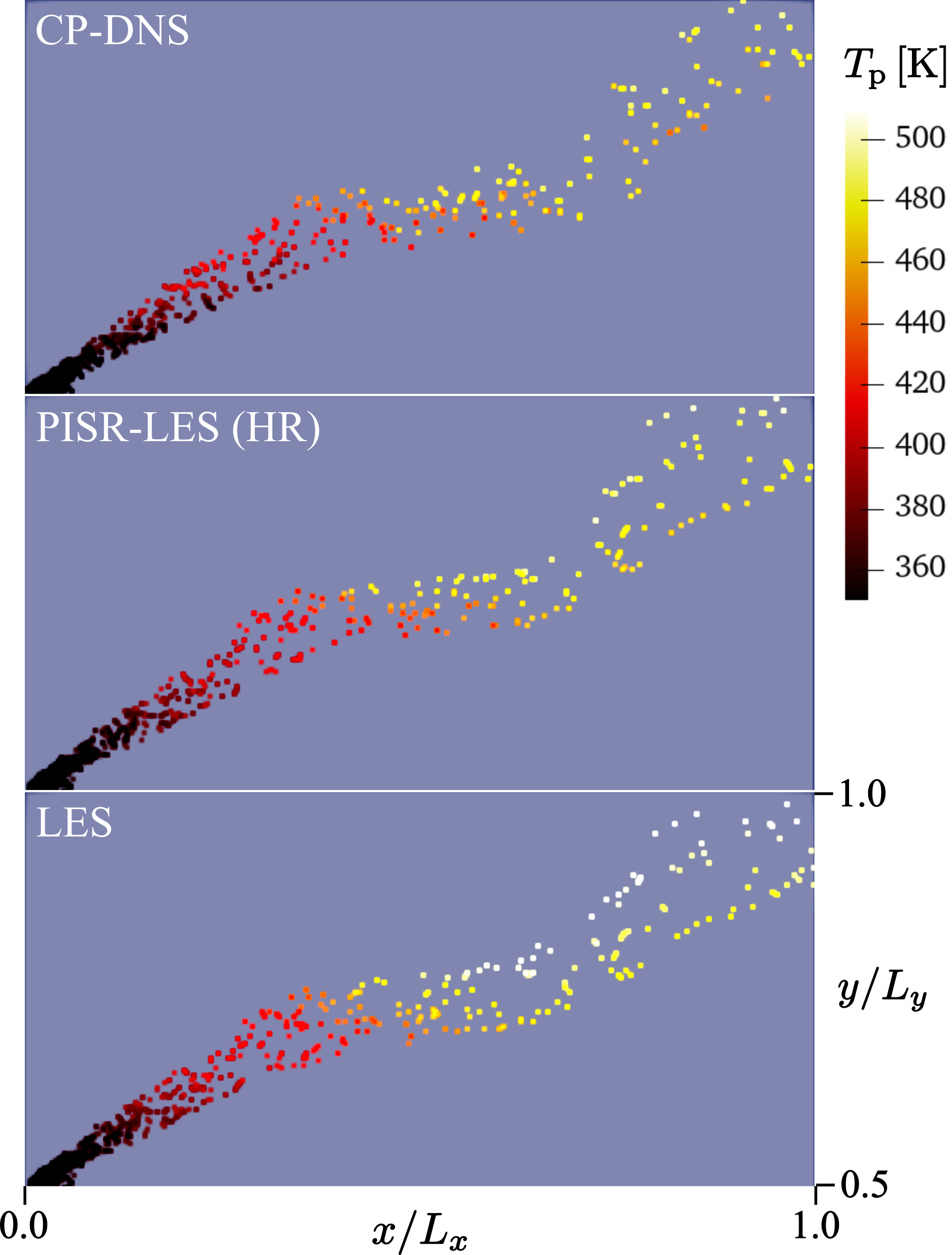}
	\vspace{-5 pt}
	\caption{\captionsize Slabs of the spray cone at $t=15.6t_{\mathrm{ft}}$, $z/L_z=0.5$, slab thickness $=\Delta x_{\mathrm{LES}}$.}
	\label{fig:gen-cone}
	\vspace{-5 pt}
\end{figure}   

Figures~\ref{fig:gen-mf}-\ref{fig:gen-YPDF} show that the evaporation rates and fuel mass fraction fields of the PISR-LES remain in much better agreement with the filtered CP-DNS than those of the standard LES for all the five cases. Figure~\ref{fig:gen-cone} illustrates the spray for case HCone. The spray shape predicted by PISR-LES on the HR mesh is generally consistent with that of CP-DNS. It can also be observed that the standard LES overestimates droplet temperatures.

\begin{figure*}[!b]
	\centering
	\includegraphics[width=\textwidth]{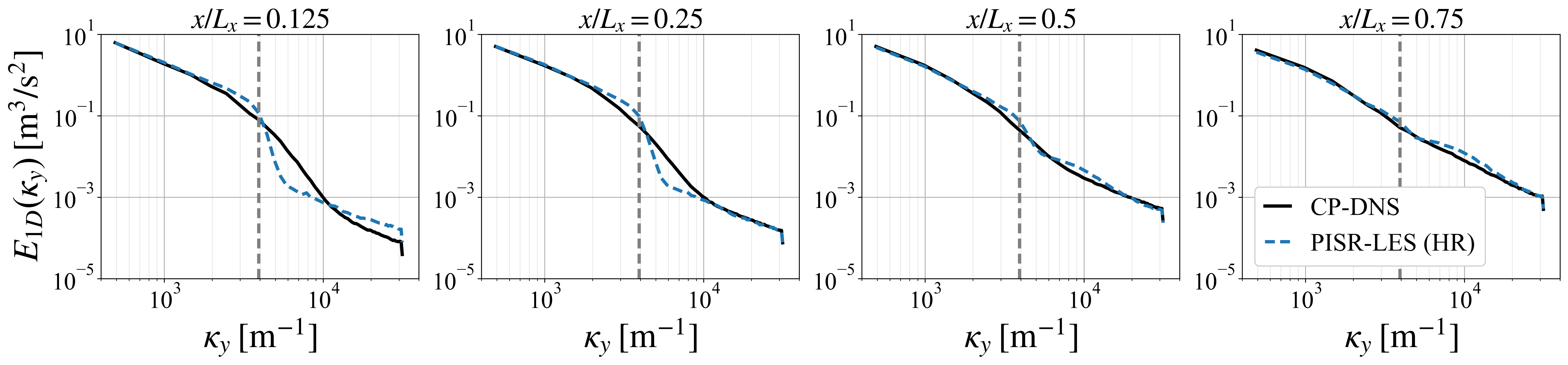}
	\caption{\captionsize One-dimensional velocity spectra of case Ret84.}
	\label{fig:gen-E1Dy}
\end{figure*}

Figure~\ref{fig:gen-E1Dy} presents the comparison of velocity spectra between the CP-DNS and the reconstructed HR fields in PISR-LES for case Ret84. A marked deviation is observed in the spectra at $x/L_x=0.125$. However, the deviation gradually narrows downstream and for $x/L_x \geq 0.5$ the agreement between the two lines is as good as that for the baseline case (Fig.~\ref{fig:apost-E1Dy}). Note that the Stokes numbers ($\mathrm{St}_{\eta}$, ratio of particle relaxation time to Kolmogorov time scale) at the inlet are very different for the two cases (cf. Table~\ref{tab:inlet},~\ref{tab:gCases}). This could be the reason for the deviation in velocity spectra in the upstream region ($x/L_x=0.125$) in Fig.~\ref{fig:gen-E1Dy}, because $\mathrm{St}_{\eta}$ is a key parameter governing the pattern of drag-induced interaction between turbulence and particles~\cite{Ferrante2003PoF,Abdelsamie2012PoF}. 
As explained in Sec.~\ref{subsec:SR}, the advantage of the PISR model in turbulence SR comes from its ability to learn the interactions between turbulence and particles in training and leverage them at inference. Therefore, its SR performance might degrade when the interaction pattern changes at inference. The recovery of model performance downstream is intriguing. Although the turbulence is decaying, the TKE of case Ret84 is still more than four times that of the baseline case at $x/L_x = 0.5$, and the $\mathrm{St}_{\eta}$ difference between the two cases remains comparable to its value at the inlet. It is inferred that the momentum interaction between the two phases downstream is dominated by the evaporation instead of drag, which exhibits similar behaviors in the two cases. As evidenced by the CP-DNS results in Figs.~\ref{fig:apost-E1Dy} and \ref{fig:gen-E1Dy}, the turbulence downstream is modulated in a similar pattern for the two cases: the velocity spectra gradually flatten and transition to a power-law scaling. Since the evaporation is modeled well in the PISR-LES (Figs.~\ref{fig:gen-mf}-\ref{fig:gen-YPDF}), the turbulence modulation by the evaporation can be effectively reconstructed. 

It should be noted that, while the present study intentionally challenges the model’s generalization capability using a limited training database, extending the database is recommended to achieve optimal performance for significantly different flow conditions.
\subsection{Computational performance\label{subsec:computational}} \addvspace{10pt}
The computational performance of CP-DNS, standard LES and PISR-LES is compared in Tables~\ref{tab:cpuhour} and~\ref{tab:mem}. The memory usage is reported for completeness. All measurements are conducted on the same computational node equipped with an NVIDIA RTX 4090 GPU for super-resolution. Table~\ref{tab:cpuhour} shows that the computational efficiency of PISR-LES, with respect to CP-DNS, is comparable to that of LES. The higher CPU hours of PISR-LES relative to LES arises from the extra super-resolution steps and advancing particles on a finer mesh. In fact, Lagrangian advancement is the most time-consuming step for both LES and PISR-LES, accounting for about 88\% of the total computing time in LES and 69\% in PISR-LES for the baseline case. As a result, the computing time of PISR-LES can be closer to that of LES when the number of tracked particles increases. This is apparent in the simulations of case D40 (cf. Table~\ref{tab:cpuhour}), where the number of particles is increased to maintain the constant $\alpha_0$.
\begin{table}[!h] \captionsize
	\caption{CPU hours$^\dagger$ for $40t_{\mathrm{ft}}$.}
	\centerline{\begin{tabular}{cccc}
			\hline 
			\rule{0pt}{1em}
			Case & CP-DNS & LES & PISR-LES     \\
			\hline
			\rule{0pt}{1em}
			Baseline  & 421 & 0.5 & 2.6          \\ 
			\rule{0pt}{1em}
			D40  & 428 & 1.2 & 3.4          \\
			\hline
			\multicolumn{4}{l}{\footnotesize $\dagger$: Wall-clock time $\times$ Number of CPU cores used}\\
	\end{tabular}}
	\label{tab:cpuhour}
	\vspace{-10 pt}
\end{table}

\begin{table}[!h] \captionsize
	\caption{Memory usage (GB) for baseline case.}
	\centerline{\begin{tabular}{cccc}
			\hline 
			\rule{0pt}{1em}
			 & CP-DNS & LES & PISR-LES     \\
			\hline
			\rule{0pt}{1em}
			CPU RAM  & 6.7 & 0.2 & 3.6         \\ 
			\rule{0pt}{1em}
			GPU VRAM  & N/A & N/A & 2.5          \\
			\hline
	\end{tabular}}
	\label{tab:mem}
	\vspace{-5 pt}
\end{table}

\section{Conclusions\label{sec:conclu}} \addvspace{10pt}
An advanced LES framework with a novel super-resolution approach -- particle-informed super-resolution (PISR) -- is proposed to improve LES of turbulent spray combustion. Simulations of moderately dense evaporating sprays show that the PISR-LES largely reproduces the gas fields at CP-DNS resolution, and thus remarkably narrows the discrepancies in the evaporation rates and fuel mass fraction fields between LES and filtered CP-DNS. The generalization tests further demonstrate the robustness of the PISR model. Importantly, the computational efficiency of PISR-LES remains comparable to that of standard LES. Note that PISR-LES is not intended exclusively for the evaporation models based on single droplet assumption. The key advantage of PISR-LES is its capability of providing high-resolution gas fields for Lagrangian advancement, which can support any evaporation model that relies on flow field details. It can also benefit any Euler-Lagrange simulations where LES resolution is insufficient for accurate particle source term modeling.  

Although the main motivation for developing PISR is to have better modeling of particle source terms, the good agreement in velocity spectra observed in the \textit{a posteriori} and generalization tests suggests the potential of PISR for turbulence closure. This potential will be explored in future work. The PISR-LES will also be extended to reacting sprays.

\section*{CRediT authorship contribution statement} \addvspace{10pt}

{\bf Ruyue Cheng}: Conceptualization, Methodology, Software, Validation, Formal analysis, Investigation, Writing-original draft, Funding acquisition. {\bf Ali Shamooni}: Software, Validation, Formal analysis, Writing-review \& editing. {\bf Andreas Kronenburg}: Conceptualization, Writing-review \& editing, Supervision, Project administration, Funding acquisition. {\bf Jan Wilhelm Gärtner}: Software, Validation, Writing-review \& editing. {\bf Thorsten Zirwes}: Formal analysis, Writing-review \& editing. 

\section*{Declaration of competing interest} \addvspace{10pt}

The authors declare that they have no known competing financial interests or personal relationships that could have appeared to influence the work reported in this paper.

\section*{Acknowledgments} \addvspace{10pt}

The authors acknowledge the financial support from the China Scholarship Council (CSC) (Grant No. 202206020071) and the German Research Foundation (DFG) project (No. 513858356).

\small
\baselineskip 10pt

\bibliographystyle{ieeetr}
\bibliography{manuscript_arXiv}

\end{document}